\documentclass[12pt]{iopart}

\usepackage{iopams}
\usepackage{graphicx}
\usepackage{iopams}

\usepackage{soul}
\usepackage{color}

   \usepackage{caption}
    \usepackage{subfigure}
    \usepackage{graphicx}
    \usepackage{bbm}
    \usepackage{geometry}
    \usepackage{comment}
\def\ket#1{\vert #1\rangle}
\def\bra#1{\langle #1\vert}

\def\3ph{\frac{3\pi}{2}}
\def\1{\mathchoice{\rm 1\mskip-4.2mu l}{\rm 1\mskip-4.2mu l}{\rm
        1\mskip-4.6mu l}{\rm 1\mskip-5.2mu l}}
\begin{document}
\title[Improvement of the polarized neutron interferometer setup demonstrating violation of a Bell-like inequality]{Improvement of the polarized neutron interferometer setup demonstrating violation of a Bell-like inequality}
\author{H Geppert\dag\ , T Denkmayr\dag\ , S Sponar\dag\ , H Lemmel\dag \ddag\ , and Y Hasegawa\dag\ }
\address{\dag\ Atominstitut, Vienna University of Technology,
Stadionallee 2, 1020 Vienna, Austria\\ \ddag\ Institut Laue Langevin, 38000 Grenoble, France\\}
\ead{hgeppert@ati.ac.at}
\begin{abstract}
For precise measurements with polarized neutrons high efficient spin-manipulation is required. We developed several neutron optical elements suitable for a new sophisticated setup, i.e., DC spin-turners and Larmor-accelerators which diminish thermal disturbances and depolarisation considerably. The gain in performance is exploited demonstrating violation of a Bell-like inequality for a spin-path entangled single-neutron state.  The obtained value of $S=2.365(13)$, which is much higher than previous measurements by neutron interferometry, is $28 \, \sigma$ above the limit of $S=2$ predicted by contextual hidden variable theories. The new setup is more flexible referring to state preparation and analysis, therefore new, more precise measurements can be carried out.

\end{abstract}
\submitto{\NJP}
\maketitle
\tableofcontents
\section{Introduction}
\label{sec:Intro}
Perfect crystal neutron interferometry was first demonstrated in 1974 at the $250$\,kW Triga MARK-II reactor in Vienna \cite{Rauch74}. Ever since neutron optical experiments, based on interference of matter waves, have provided a power full means of demonstrating effects related to fundamental aspects of quantum physics \cite{Rauch00}, such as measuring the $4\pi$-periodicity of fermions \cite{Rauch75}, 
gravitational effects on the neutron \cite{Werner75}, spin superposition \cite{Summhammer83, Badurek83} and topological phases \cite{Wagh97, Hasegawa96, Filipp05}. Entanglement between different degrees of freedom like the neutron's spin, energy and path have been accomplished \cite{Hasegawa10} and used for testing Bell's inequality \cite{Hasegawa03, Erdosi11} or measuring the influence of geometric phases \cite{Sponar10}. Such entanglement is achieved within single particles. Further demonstrations of the contextual nature of quantum mechanics (QM) have been performed successfully using neutron interferometry \cite{Hasegawa06, Bartosik09, Denkmayr13}.

The violation of the Bell inequality can only be shown with high interference contrast and high spin polarisation. In the first experiment \cite{Hasegawa03} a Mu-metal sheet was used as a spin turner, which induced dephasing due to small angle scattering and thereby reduced the interference contrast. The next setup \cite{Erdosi11} solved the problem of dephasing but the degree of polarisation became problematic.

In this paper we report a significantly improved experimental setup. We designed new DC spin-turners and Larmor-accelerators which allow for very high contrast of the interference fringes and high temperature stability during long measurements. They also enable high degrees of polarisation and high efficiency spin manipulation. This setup allows a large variety of state preparations and therefore provides capability for many future experiments \cite{Denkmayr13, Sponar14}. We performed a test of Bell's inequality using this new setup. The results reveal the substantial improvements achieved by the newly designed setup. 
\section{Improvement of the polarised interferometer setup}
\label{sec:Improvement of the polarised interferometer setup}
\subsection{Overview of the polarised interferometer setup}
\begin{figure}
 	 \begin{center}
 	 \scalebox{1}{\includegraphics{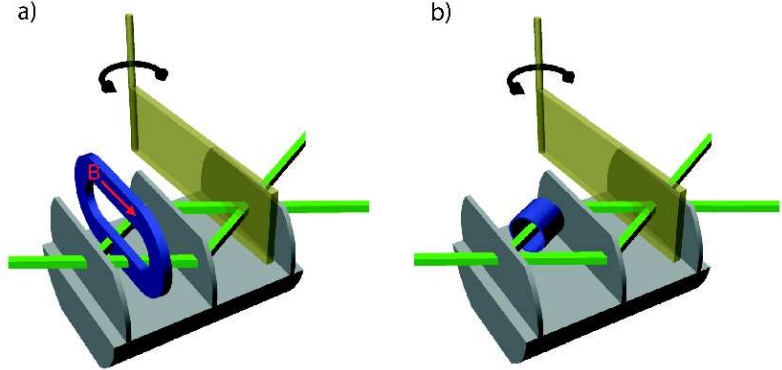}} 
 	 \end{center}
 	 \caption{a) setup with mu-metal inserted in the beam to turn the spin causing dephasing. b) setup using a mu-metal ring to turn the spin causing depolarisation due to inhomogeneity of the field.}
 	 \label{fig:old_setup}
\end{figure}

In our setup high degrees of polarisation, thermal stability, efficient spin-manipulation and spin-analysis are required. Former setups had drawbacks that degrade the quality of the measurement results. These setups used conventional DC spin-turners or RF-flippers inside the IFM. Such setups were used for spin-superposition, geometric phase and entanglement measurements \cite{Summhammer83, Sponar10, Bartosik09}. For earlier Bell-measurement using single-neutron interferometry two different setups were realized \cite{Hasegawa03, Erdosi11}. In both setups the spin manipulation in the interferometer (IFM) was problematic: The contrast and the degree of polarisation were reduced. These two setups are shown in figure \ref{fig:old_setup}.
Scheme \ref{fig:old_setup}a) shows the IFM with inserted soft magnetic Mu-metal foil as a spin turner. This is achieved by a magnetic field induced into the Mu-metal by a DC-coil outside of the IFM. The Mu-metal foil considerably reduced the contrast of the IFM due to dephasing. To overcome this problem  another setup was designed, which does not need any material in the neutron beam in the IFM \cite{Erdosi11}, shown in figure \ref{fig:old_setup}b). In one path of the IFM the beam passes a tube of Mu-metal which reduces the strength of the magnetic guide field and thereby induces a relative spin rotation by different Larmor precession in the two IFM paths. Since the guide field leaks into the cylinder at its open ends, the field homogeneity is compromised which causes depolarisation of the neutron beam. This setup also requires a spin turner in front of the IFM which additionally reduces the degree of polarisation as described below. 
\begin{figure}
 	 \begin{center}
 	 \scalebox{1}{\includegraphics{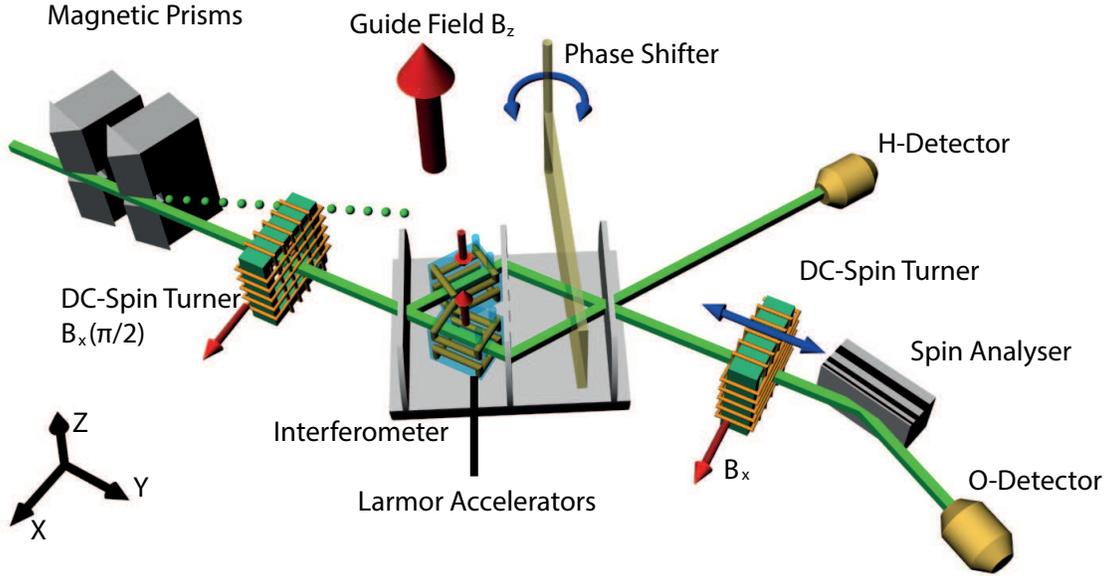}} 
 	 \end{center}
 	 \caption{Experimantal setup for measuring Bell inequalities using a triple Laue interferometer. Magnetic prisms are used to polarize the incoming beam. To avoid depolarisation a magnetic guide field $B_z$ is applied around the hole setup. A spin rotator before the IFM turns the spin into the xy-plane. The first plate of the interferometer splits the beam. In each path a Larmor accelerator turns the spin by $\pm \pi/2$ respectively. With a phase shifter the relative phase $\chi$ can be tuned. The two exit beams are monitored by the O- and H-detectors. The beam arriving at the O-detector is filtered by a spin analyzer.}
 		 \label{fig:Setup}
\end{figure}
A schematic view of the new setup is shown in figure \ref{fig:Setup}. The beam is monochromatised to have a mean wave length of $\lambda_0 = 1.92(2)$~\AA  \,by a silicon channel-cut perfect-crystal monochromator. The incoming neutron beam is polarized by two birefringent magnetic prisms which deflect beams of  up- and down-spin neutrons in different directions. The angle between these two beams is $2.3\, 10^{-5}$ \,rad. Since the acceptance width of the interferometer crystal for Laue diffraction is even smaller, we can select one of the spin components (spin-up) by adjusting the rotation angle of the IFM accordingly. Neutrons with spin-down pass the IFM without being reflected and are blocked by a beam stopper afterwards. To avoid depolarisation of the beam a guide field is applied over the entire setup.
In front of the IFM the spin is rotated by a DC spin-turner into the xy-plane. Within this plane we can adjust the spin by utilizing Larmor precession without putting any material into the beam. This is important to avoid loss of interference contrast due to dephasing. A sapphire phase shifter of $5\,$mm thickness  between second and third plate of the IFM tunes the relative phase $\chi$ between the beams in path I and path II. Behind the IFM the spin analysis is carried out using  a DC-coil on a translation stage together with a Co-Ti super-mirror array. The neutrons are detected in $^3$He counters with more than 99\% efficiency \cite{Bergmann}.

\subsection{$\pi /2$-Spin rotator}
\begin{figure}
 	 \begin{center}
 	 \scalebox{0.9}{\includegraphics{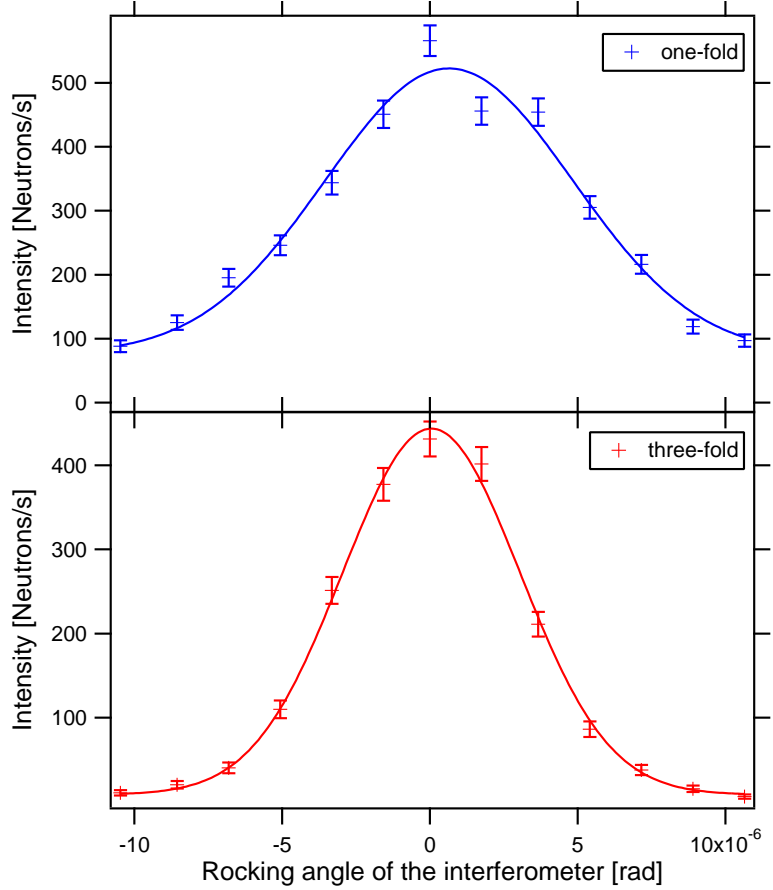}}
 	 \end{center}
 	 \caption{Rocking curves for one- and three-fold reflecting monochromator crystal. The FWHM's are $6.11(47) \, 10^{-6}$ rad for the 1-fold and $4.26(10) \, 10^{-6}$ rad 3-fold monochromator crystals.}
 	 \label{fig:Crystal}
\end{figure}

The $\pi/2$-spin rotator is placed between the magnetic prisms and the IFM. Due to the small separation of spin-up- and spin-down -beam by the magnetic prisms and the fact that the selection of the peak takes place at the first plate of the IFM, wider peaks of the IMF's rocking curves degrade the degree of polarization of the neutron beam. The peak width at the first IFM plate is determined by the monochromator and the properties of the $\pi/2$-spin rotator regarding small-angle scattering. In contrast to earlier experiments where a single reflecting monochromator was used, we used a three-fold channel-cut monochromator. A comparison of rocking curves by using one- and three-fold reflection is shown in figure \ref{fig:Crystal}.

The full width at half maximum $\sigma$ (FWHM) of the rocking peak of the single reflecting crystal has a FWHM$ = 6.11(47) \, 10^{-6}$ rad, whereas the triple reflecting crystal has a FWHM$ = 4.26(10) \, 10^{-6}$ rad: the former is 70\% wider than the triple reflecting crystal.

\begin{figure}
 	 \begin{center}
 	 \scalebox{0.9}{\includegraphics{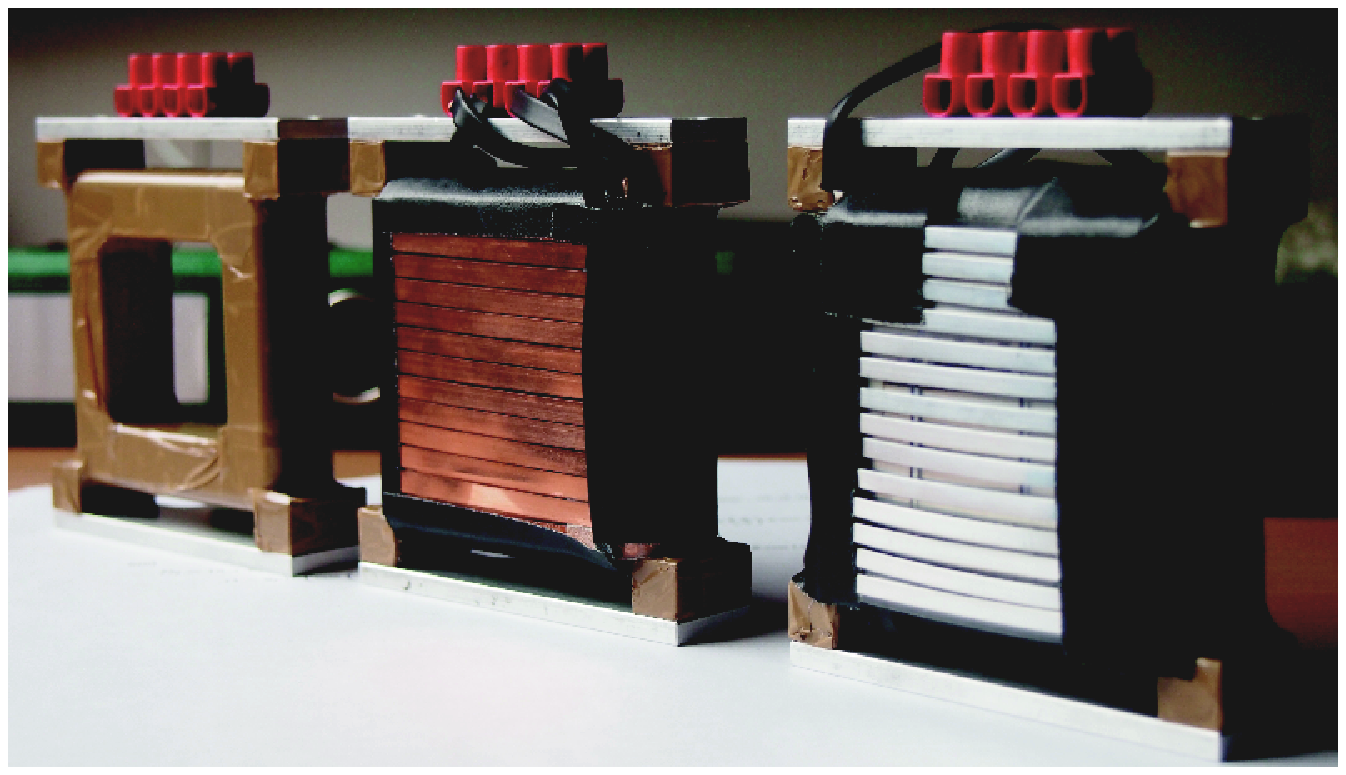}} 
 	 \end{center}
 	 \caption{ Photographs of $\pi/2$-spin turner coils made out of aluminium, copper and an empty frame for a coil (right to left).}
 	 \label{fig:DC_Coils}
\end{figure}

\begin{figure}
 	 \begin{center}
 	 \scalebox{0.9}{\includegraphics{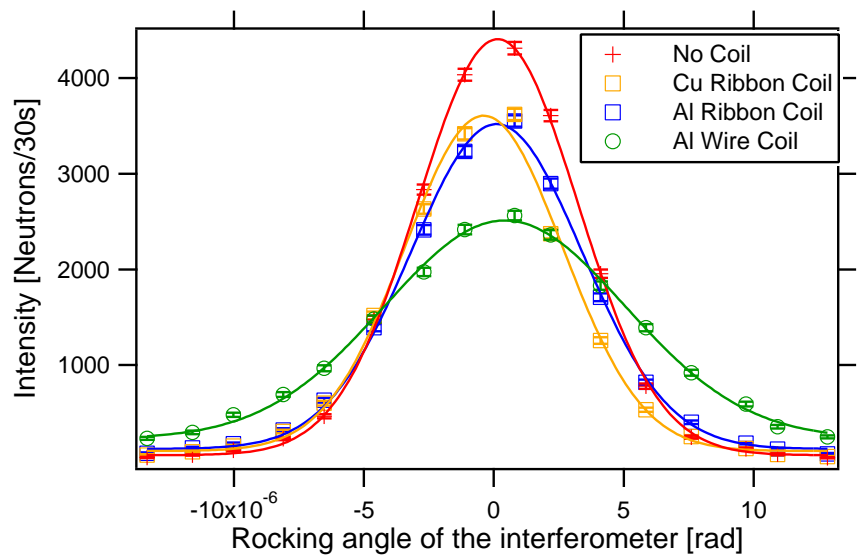}} 
 	 \end{center}
 	 \caption{Rocking curves without coil, with the copper ribbon coil, the aluminium the ribbon coil and the aluminium wire coil inserted in the neutron beam.}
 	 \label{fig:Peakcomp}
\end{figure}

\begin{table}
\begin{center}
\caption{Rocking curve comparison with spin turning coils made of different materials, normalised to the empty setup.}
\begin{tabular}{l@{\hskip 10mm} c c}
\hline
& {Peak} &{FWHM} \\
\hline
No Coil & $1.000$&$ 1.000$ \\ 
Al Wire &$0.56(1)$&$ 1.68(4)$ \\ 
Al Ribbon &$0.80(1)$&$1.16(2)$ \\ 
Cu Ribbon 3mm width&$0.84(1)$&$1.11(2)$ \\
Cu Ribbon 4mm width&$0.85(1)$&$1.16(2)$ \\ \hline
\end{tabular}
\label{tab:coilcompaire}
\end{center}
\end{table}

\begin{figure}
 	 \begin{center}
 	 \scalebox{0.9}{\includegraphics{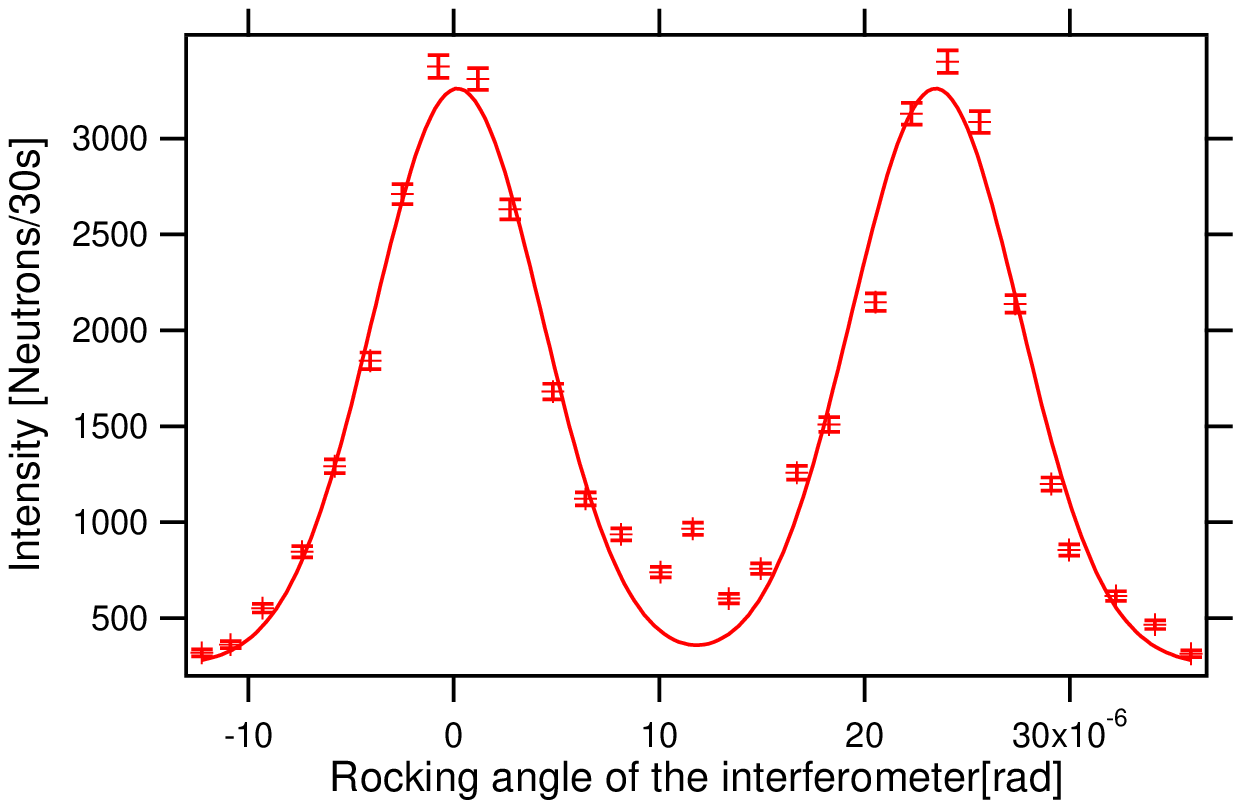}} 
 	 \end{center}
 	 \caption{Rocking curve showing the up-spin and the down-spin peak produced by the birefringent magnetic prisms.}
 	 \label{fig:double_peak}
\end{figure}

The peak broadening was measured for Different coils made of copper ribbon ($0.1 \times 3\,$mm$^2$ and $0.1 \times 4\,$mm$^2$ in profile), aluminium ribbon ($1 \times 4\,$mm$^2$ in profile), and aluminium wire (0.5\,mm in diameter). In figure \ref{fig:DC_Coils} the aluminium ribbon coil and the copper ribbon coil are shown next to an empty coil frame. Figure \ref{fig:Peakcomp} shows the rocking curves for different coils in comparison to the empty. In table \ref{tab:coilcompaire} the peak height and the width of the rocking curves for different coils are given in respect to the empty beam line. One can see that despite the small absorption and scattering cross-section of aluminium, the wire coil enlarges the width of peak and lowers the peak intensity because of small-angular scattering, whereas within the error there is no evident difference between the three ribbon coils. The rectangular profiled ribbons do not produce significant small-angle scattering and therefore more neutrons fulfill Bragg's law. For our measurements we used $3\,$mm wide copper ribbon for both DC spin-turners. In figure \ref{fig:double_peak} one can see the separation of the up-spin peak and the down-spin peak with the $\pi/2$-spin rotator in the beam line. Between the two main peaks one can see a small peak produced by neutrons that fulfil higher orders of the Bragg condition. Applying the two-flipper method by tuning two spin rotators before and after the IFM the degree of polarisation was measured to be $>0.993$, $0.98(1)$ for the efficiency of the first coil and $0.98(1)$ for the efficiency of the second coil. The efficiency of the $\pi/2$-spin rotators is reduced by the stray fields of the magnetic prisms in front of the IFM and of the super-mirror spin analyser behind it respectively. Therefore the coils were placed as far from super mirror and magnetic prisms as possible.
\subsection{Larmor-Spinrotator}
\begin{figure}[ht]
 	 \begin{center}
 	 \scalebox{0.6}{\includegraphics{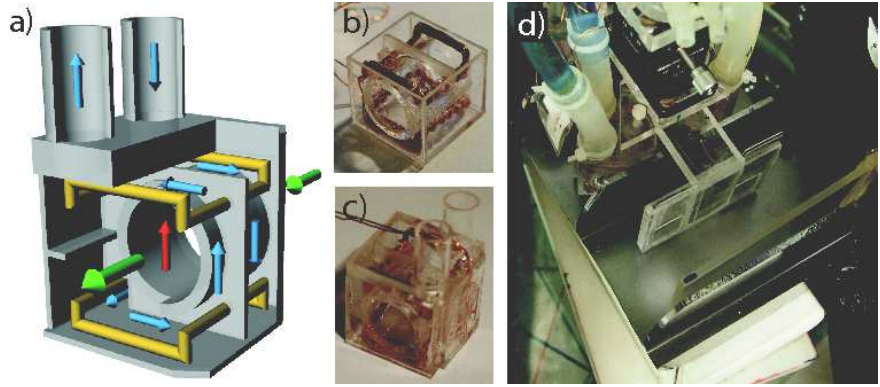}} 
 	 \end{center}
 	 \caption{a) schematic view of the Larmor accelerator boxes with coils in Helmholtz geometry (yellow), magnetic field (red), neutron beam (green) and water flux (blue), b) Larmor acceleration coil with Helmholtz geometry in box,
 	 c) closed box with connectors for water cooling,
 	 d) boxes in the interferometer with connected cooling system and absorber holder for adjustment of the Larmor accelerators.}
 	 \label{fig:Fotos_box}
\end{figure}
The state preparation requires two Larmor accelerator coils placed in the IFM as shown in figure \ref{fig:Setup}. These coils in Helmholtz geometry apply an additional parallel or anti-parallel field to the guide field in z-direction locally and thereby change the spin precession in the xy-plane. Since the rotation angle is given by $\alpha (B_z)= \frac{2 \mu l}{\hbar v} B_z$, where $\mu$ is the magnetic moment of the neutron, $l$ is the length of the coils and $v$ is the velocity of the neutrons, a magnetic field of about $0.33$\, mT is required for a spin rotation of $\pi /2$. For the fabricated coil a current of about $0.7$\,A is required for a spin rotation of $\pi /2$. Since the coils produce heat and due to the high sensitivity of the IFM to thermal influences, the coils need to be cooled down. To obtain constant temperature up to $0.1$\,$^ \circ$C the coils are placed in  small boxes which are completely flooded with temperature controlled water. The coil wire is in direct contact with the water, insulated only by lacquer. The boxes are made of acrylic glass, which is a thermal and electrical insulator. Length, width and height of the boxes amount to 22, 26 and 26 mm respectively. Figure \ref{fig:Fotos_box}a) shows a schematic view of a box. The boxes have a  straight passage for the neutron beam (green arrows), so that the beam doesn't pass any material and therefore no dephasing occurs. The figure also shows the magnetic field in z-direction (red arrow) applied by the coils in Helmholtz configuration (yellow). The flux of the cooling water is depicted by blue arrows. Figures \ref{fig:Fotos_box}b) and \ref{fig:Fotos_box}c) show the box without top and finished with the connectors for the water cooling respectively. In figure \ref{fig:Fotos_box}d) one can see the boxes placed in the IFM. In this picture the mountings for beam stoppers between second and third plate of the IFM are depicted. The beam stoppers are used to calibrate the Larmor accelerators one at a time by blocking the other path respectively. The beam stoppers used here are $1$\, mm thick cadmium plates.

\section{Violation of Bell-like inequalities}
 \label{sec:Violation of Bell-like inequalities}
Quantum mechanics (QM) is one of the most successful physical theories and its predictions have been proven accurately in many experiments using various kinds of systems. Einstein, Podolsky and Rosen (EPR) \cite{EPR35} argued that QM is not a complete theory since it only gives probabilistic predictions and that there must be an underlying deterministic theory to QM.  In 1964 Bell \cite{Bell64} showed that local hidden variable theories satisfy some inequalities that are violated by QM. Shortly after Bell published his well known paper Clauser, Horne, Shimony, and Holt (CSHS) reformulated Bell's inequalities suitable for the first experimental test of quantum non-locality \cite{Clauser69, Clauser78}. In the case of neutrons not two particles are entangled but two different degrees of freedom within one particle \cite{Mermin90, Mermin93}. Non contextual hidden variables theories (NCHVT) states that the outcome of a measurement is independent of previous or simultaneous measurements on any set of commuting observables.
\subsection{Theory}
In our single neutron interferometer we entangle two different degrees of freedom (spatial and spin) of a single neutron \cite{Basu01}. The neutron is described by a tensor product Hilbert space $H = H_P \otimes H_S$, where $H_P$ corresponds to the spatial wave function and $H_S$ to the spinor wave function. Since observables of the spacial part commute with those of the spinor part one can derive a Bell-like state. The normalized wave function is given by
\begin{equation}\label{eq:instate}
\hspace{-10mm}
{\ket\Psi =\frac{1}{\sqrt{2}} \big( \ket \uparrow \otimes \ket I + \ket \downarrow \otimes \ket{II} \big)}.
\end{equation}
Where $\ket\uparrow$ and $\ket\downarrow$ correspond to up- and down-spin and $\ket I$ and $\ket{II}$ represent the two paths in the IFM. The expectation value of the joint spin and path measurement can be written as 
\begin{equation}\label{eq:expdeff}
\hspace{-10mm}
E(\alpha,\chi)= \bra \psi \hat{P}^S (\alpha) \cdot \hat{P}^P (\chi) \ket \psi = \bra \psi [ \hat{P}^S _{\alpha ; 1} -  \hat{P}^S _{\alpha ; -1}] \times [ \hat{P}^P _{\chi ; 1} - \hat{P}^P _{\chi ; -1} ] \ket \psi.
\end{equation}
The observables for spin $\hat{P}^S (\alpha)$ and path $\hat{P}^P (\chi)$ can be decomposed by projection operators $ \hat{P}^S _{\alpha ; \pm 1}$ and $\hat{P}^P _{\chi ; \pm 1}$, which project onto orthogonal spin states $ \frac{1}{\sqrt{2}}  ( \ket \uparrow \pm e^{\rmi \alpha} \ket \downarrow ) $ and orthogonal path states $ \frac{1}{\sqrt{2}}  ( \ket I \pm e^{\rmi \chi} \ket {II} ) $, respectively. The expectation value given in equation \ref{eq:expdeff} and the projection operators correspond to $ P'_\pm (a)$ , $P'_\pm (b)$ and $E'(a,b)$ in the conventional EPR argument \cite{Aspect84}. In the experiment the parameter $\alpha$ can be varied by polarisation measurement of the Bell-like state, $\chi$ is tuned by an auxiliary phase shifter.
For single-neutron interferometry a Bell-like inequality can be expressed using the expectation values $E(\alpha , \chi )$ as $ -2 \leq S \leq 2$, with
\begin{equation}\label{eq:Svalue}
\hspace{-10mm}
{S \equiv E(\alpha_1,\chi_1)+E(\alpha_1,\chi_2)-E(\alpha_2,\chi_1)+E(\alpha_2,\chi_2)}.
\end{equation}
In our experiment the expectation values $E(\alpha,\chi)$ are determined by a combination of count rates $N(\alpha,\chi)$ of a single detector donated to appropriated settings of $\alpha$ and $\chi$. This gives:
\begin{equation}
\label{eq:expval}
\hspace{-10mm}
{E( \alpha , \chi )=\frac{N(\alpha , \chi)+N(\alpha + \pi, \chi + \pi)- N(\alpha , \chi + \pi)- N(\alpha + \pi , \chi)}{N(\alpha , \chi)+N(\alpha + \pi, \chi + \pi)+ N(\alpha , \chi + \pi)+ N(\alpha + \pi , \chi)}}
\end{equation}
The count rates $N(\alpha, \chi )$ are given by $N(\alpha,\chi)= \frac{1}{2}[1-\cos (\alpha+
\chi)]$ according to quantum mechanical predictions. This leads to a sinusoidal behaviour of the expectation values $E(\alpha,\chi)=\cos(\alpha + \chi)$. Bell's inequality are violated for various set of polarisation analysis $(\alpha)$ and phase shifts $(\chi)$, but the largest violation is expected for $\alpha_1= 0$, $\alpha_2=\pi /2$, $\chi_1=\pi/4$ and $\chi_2= -\pi/4 $, which gives the value $S= 2 \sqrt{2}= 2.82 > 2$. By measuring $S$ one can test whether or not NCHVTs can describe nature correctly.
\subsection{Measurement}
\begin{figure}[ht]
 	 \begin{center}
 	 \scalebox{0.6}{\includegraphics{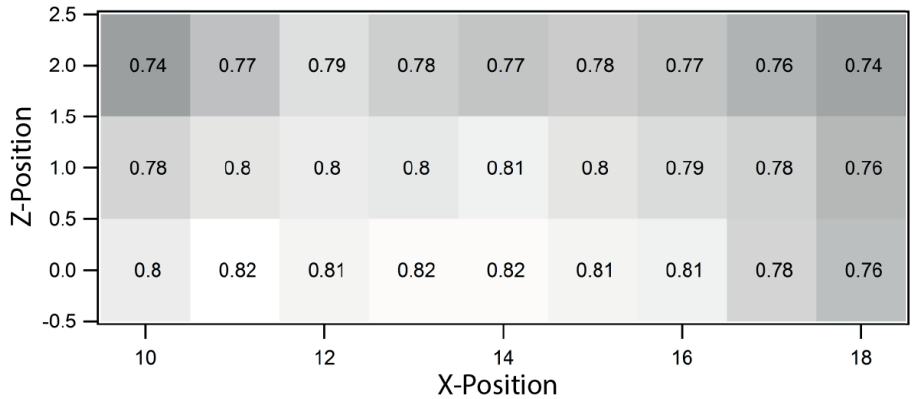}} 
 	 \end{center}
 	 \caption{Raster scan showing the contrast of the IFM in respect to position in xz-plane.}
 	 \label{fig:rasterscan}
\end{figure}

\begin{figure}[ht]
 	 \begin{center}
 	 \scalebox{0.8}{\includegraphics{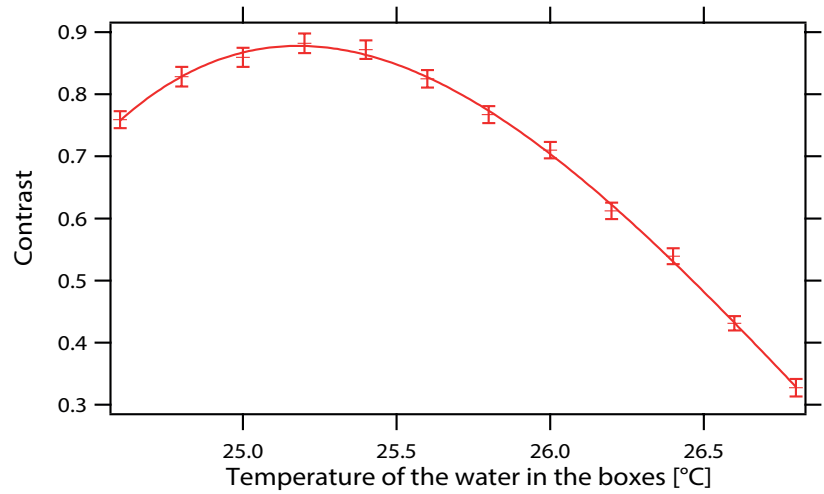}} 
 	 \end{center}
 	 \caption{The contrast of the interferometer as a function of temperature. At 25.2$^\circ$C a contrast $C>0.88$ can be achieved, at 1$^\circ$C higher temperature only $C<0.60$ can be reached.}
 	 \label{fig:CvsT}
\end{figure}

\begin{figure}[ht]
 	 \begin{center}
 	 \scalebox{0.9}{\includegraphics{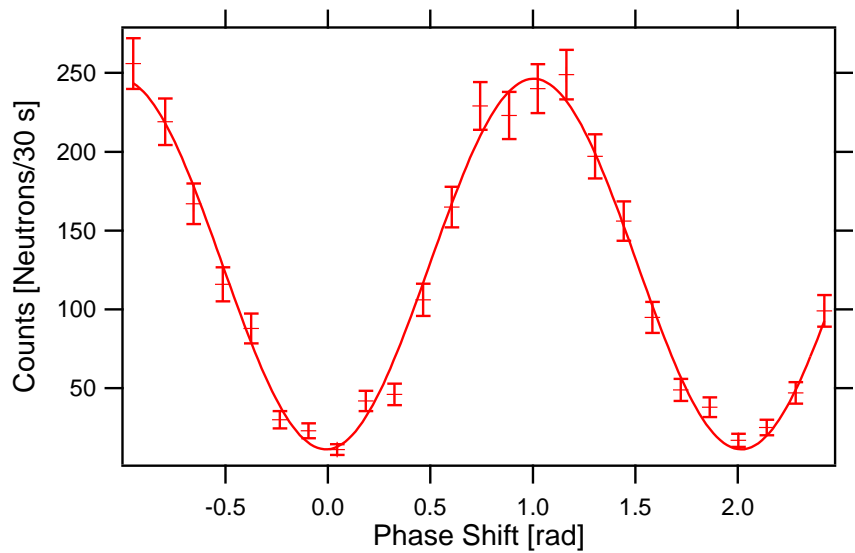}} 
 	 \end{center}
 	 \caption{Interference pattern with a contrast of 0.91.}
 	 \label{fig:high_contrast}
\end{figure}

For optimization of a measurement every part of the setup should be adjusted in a systematic way. First of all the point of highest interference contrast on the IFM, i.e. the sweet spot is looked for. The interferometer crystal is mapped by moving an aperture vertically and horizontally in the xz-plane in front of the IFM and measuring the contrast on each position. The result of such a raster scan is shown in figure \ref{fig:rasterscan}. This scan is performed with a beam cross-section of $3\times3$\,mm$^2$ and a step increment of $1$\,mm. One can see that only a small part of the IFM provides highest contrast, which  reaches at $C=0.82$. 
The contrast of the IFM is very sensitive to temperature fluctuations. Even temperature changes by  $0.1$\,$ ^\circ $C destroy the interference pattern. Since the guide field and the Larmor-accelerators produce heat, both elements are water cooled. This is done by two temperature stabilized water pumps. To optimize the temperature of the cooling water for the guide field and the boxes, temperature scans are performed.
In figure \ref{fig:CvsT} the contrast of the IFM is plotted for different temperatures of the cooling water in the boxes. For $25.2$\,$^\circ $C an average contrast of $C>0.88$ is achieved. After stabilisation contrast up to $C=0.91$ can be observed as seen in figure \ref{fig:high_contrast}. When the temperature is raised up to $26.8$\,$^\circ$C the contrast drops to $C<0.33$. A raise by $1$\,$^\circ$C in temperature results in a decrease in contrast of $C=0.60$.
\begin{figure}[ht]
 	 \begin{center}
 	 \scalebox{0.8}{\includegraphics{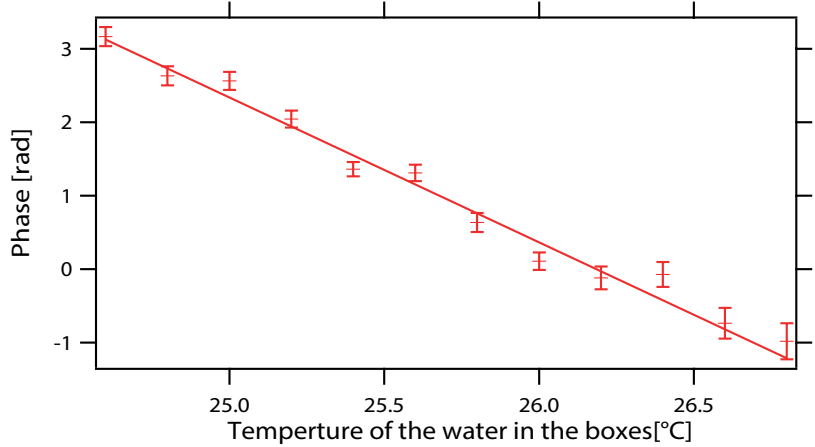}} 
 	 \end{center}
 	 \caption{The phase shift induced by temperature change of the Larmor accelerators.}
 	 \label{fig:Phase_vs_T}
\end{figure}
Thermal stability is important for another crucial point, since not only loss in contrast but also considerable phase drifts occur. A temperature change of $1$\,$^\circ$C in the boxes results in $1.92\,$rad phase shift. Figure \ref{fig:Phase_vs_T} shows this situation. The large error bars at high temperatures arises from the low contrast obtained at this temperatures as seen in figure \ref{fig:CvsT}. Both the loss in contrast as well as the phase shift resulting from thermal instability degrade the quality of measurement results. To increase the stability of the setup the IFM and the Larmor accelerators are placed in a box to avoid air convection and therefore temperature fluctuations over long periods of time.
\begin{figure}[ht]
 	 \begin{center}
 	 \scalebox{1}{\includegraphics{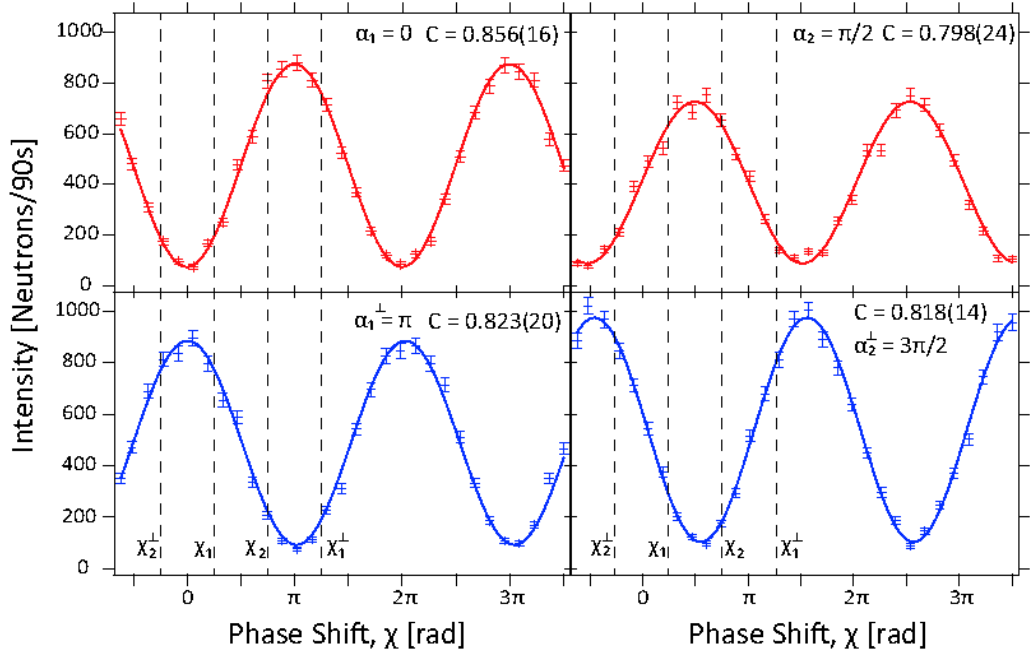}} 
 	 \end{center}
 	 \caption{Intensity oscillations obtained by tuning the phase shift $\chi$ for four different spin analysis $\alpha$ $C$, give the measured contrast.}
 	 \label{fig:Bell_complete}
\end{figure}
To determine the expectation values $E(\alpha,\chi)$ occurring in the Bell-inequalities four spin-directions $\alpha_1= 0$, $\alpha_2=\pi/2$, $\alpha^{\perp}_1=\pi$, $\alpha^{\perp}_2=3\pi/2$ and four phase-shifts $\chi_1= \pi /4$, $\chi_2= 3\pi /4$, $\chi^{\perp}_1= 5\pi /4$,  $\chi^{\perp}_2= 7 \pi /4$ need to be measured. The spin-directions are selected by the DC spin-turner behind the IFM, while the phase shifts are tuned by the sapphire phase shifter inside the IFM. A set of this measurements is shown in figure \ref{fig:Bell_complete}. The data is fitted to a sinusoidal function using least squares fit method. The error results from statistical fluctuations in count rate and systematic errors due to imperfect spin manipulation and phase instabilities during the measurement. We repeated the measurements twice, in order to reduce statistical errors.
By evaluating the Bell inequality the S-value in equality (\ref{eq:Svalue}) is calculated to be
\begin{equation}
\label{eq:result}
S=2.365 \pm 0.013,
\end{equation}
which implies a violation by more than $28 \sigma$.
\section{Discussion and Conclusion}
 \label{sec:Conclusion}
We have presented a new design of the neutron-interferometer setup and new devices for spin manipulation, which considerably improve the abilities of the polarized neutron IFM. Using a new coil design for the DC spin-turners and a channel-cut monochromator the degree of polarisation of the incoming beam of $>0.993$ is achieved. New Larmor accelerators allow the reduction of thermal disturbances on the IFM and dephasing since no material is put in the beam path inside the IFM. This enables high contrasts up to $C=0.91$. The newly designed spin manipulators allow easy and precise manipulation of the neutron's spin and enable various applications for future experiments. 
With this setup we obtained the value of $S= 2.365(13)$ for Bell-like inequality measurements, which is $28 \sigma$ above the boarder of 2 and as a consequence disproves NCHVT clearly. 
\ack
This work has been supported by the Austrian Science Fund (FWF), Project. Nr. P25795 -N02 and P24973-N20.
\newpage
{References}
\vspace{5mm}
\providecommand{\newblock}{}

\end{document}